# Relativistic Corrections to the Electromagnetic and Axial Moments of Nuclei and Other Composite Systems*

Felix Schlumpf
Center for Theoretical Physics, Department of Physics
University of Maryland, College Park, Maryland 20742

and

Stanley J. Brodsky
Stanford Linear Accelerator Center
Stanford University, Stanford, California 94309

**Abstract**

We calculate the electromagnetic and axial nuclear moments of the deuteron and triton as a function of their radius using a relativistic two-nucleon and three-nucleon model formulated on the light-cone. The results also provide an estimate of the nuclear binding corrections to helicity-dependent deep inelastic scattering sum rules. At large nucleon radius, the moments are given by the usual non-relativistic formulae modified by finite binding effects. At small radius, the moments take the canonical values given by the generalization of the Drell-Hearn-Gerasimov sum rule. In addition, as $R \to 0$, the constituent helicities become completely disoriented, and the Gamow-Teller matrix element vanishes. Thus, in the pointlike limit $MR \to 0$, the moments of a spin-one bound states coincide with the canonical couplings of elementary spin-one bosons of the Standard Model, $\mu = e/M, Q = -e/M^2$, and $g_A = 0$.

Submitted to Physics Letters B.

*This work was supported in part by the National Science Foundation and in part by the Department of Energy, contract DE-AC03-76SF00515.

Although the physical structure of spin-one nuclei, spin-one mesons, and the gauge bosons of the Standard Model are highly disparate, certain features of their electromagnetic interactions are universal, such as the ratios of their form factors $G_C(Q^2)/G_M(Q^2)$, and $G_C(Q^2)/G_Q(Q^2)$ at large momentum transfer [1]. This universality of the spin-one form factors reflects the underlying gauge and chiral symmetry of QCD at short distances. More remarkably, one can use a generalization [2] of the Drell-Hearn-Gerasimov sum rule [3] to show [1] that the magnetic and quadrupole moments of any composite spin-one system take on the canonical values $\mu = e/M$ and $Q = -e/M^2$ in the limit of zero bound-state radius or infinite excitation energy. Thus in the strong binding limit, the moments of composite particles coincide with the moments of the gauge particles in the tree-graph approximation to the Standard model. In this paper, we shall investigate the quantitative behavior of nuclear axial and electromagnetic moments for both strong and weak binding limit as well as demonstrate the transition between them.

An important feature of the relativistic treatment of the moments of composite systems is the non-additive spin structure induced by the Wigner boost. This leads to the remarkable result that one obtains a non-zero contribution to the quadrupole moment even if the deuteron has no $D-$wave contribution. The same non-additive spin structure is required to reproduce the low energy theorem for Compton scattering on a composite system as well as the Drell-Hearn-Gerasimov sum rule [3] for polarized photoabsorption cross sections [4].

The light-cone ("front-form") formalism [5] provides a convenient covariant framework for evaluating current matrix elements of composite systems [6]. The formalism is independent of the choice of momentum $p^\mu$ and form factors can be calculated from diagonal matrix elements; $i.e$, the convolution of light-cone wavefunctions with the same particle number $n$. In contrast, in equal-time theory, one needs to consider frame-dependent non-diagonal pair creation matrix elements as well as vacuum creation contributions to the current which are unconstrained by the Fock wavefunctions. The Bethe Salpeter formalism is covariant, but one needs to evaluate the matrix elements of an infinite number of irreducible kernels, even in the case when one constituent is infinitely heavy.

Our light-cone analysis is similar to the one given in Ref. [7] for the moments of the proton. Since the light-cone approach incorporates the correct relativistic properties of the interactions of composite systems, it provides a reliable method to evaluate nuclear binding corrections to nuclear moments and deep inelastic structure functions.

The light-cone model given in Ref. [8] provides a simple framework for representing the general structure of a two-body relativistic wavefunction [9, 10, 11]. In the model, the bound-state wavefunction is constructed as a product of a momentum wavefunction and a spin-isospin wavefunction, which is uniquely determined by symmetry requirements. A Wigner (Melosh) [12] rotation is applied to the spinors, so that the wavefunction of the nucleus is an eigenfunction of $J$ and $J_z$ in its rest frame [13, 14]. Since we only consider the S-wave component of the wavefunction, we choose for the momentum dependence

$$\psi(\mathcal{M}^2) = N \exp\left(\frac{-\mathcal{M}^2}{2\beta^2}\right), \qquad (1)$$

where $\beta$ sets the scale of the nuclear size. As we have shown in Ref. [7], the predictions for the electromagnetic and axial moments are essentially independent of the exact shape of $\psi(\mathcal{M}^2)$ once the physical radius, as defined from the slope of the form factor, is fixed. The invariant mass $\mathcal{M}$ can be written as

$$\mathcal{M}^2 = \sum_{i=1}^{2} \frac{\vec{k}_{\perp i}^2 + m_N^2}{x_i}, \qquad (2)$$

where we used the longitudinal light-cone momentum fractions $x_i = p_i^+/P^+$ ($P$ and $p_i$ are the deuteron and nucleon momenta, respectively, with $P^+ = P^0 + P^z$). The internal momentum variables $\vec{k}_{\perp i}$ are given by $\vec{k}_{\perp i} = \vec{p}_{\perp i} - x_i \vec{P}_\perp$ with the constraints $\sum \vec{k}_{\perp i} = 0$ and $\sum x_i = 1$. The Melosh rotation has the matrix representation [15]

$$R_M(x_i, k_{\perp i}, m_N) = \frac{m_N + x_i \mathcal{M} - i\vec{\sigma} \cdot (\vec{n} \times \vec{k}_i)}{\sqrt{(m_N + x_i \mathcal{M})^2 + \vec{k}_{\perp i}^2}}, \qquad (3)$$

with $\vec{n} = (0, 0, 1)$, and it becomes the unit matrix if the quarks are collinear $R_M(x_i, 0, m) = 1$. Thus the internal transverse momentum dependence of the light-cone wavefunctions also affects its helicity structure.

The parameter $\beta$ for the deuteron has been chosen to give approximately the same $S$-state wavefunction as the one given in Ref. [16]; namely $\beta_d = 0.12$ GeV. Since the binding energy for the triton is larger then the one for deuterium, we also use a larger $\beta$; namely $\beta_t = 0.28$ GeV.

We can evaluate the three form factors of a spin-one system in terms of the light-cone spin matrix elements of the plus component of the current [6, 1, 13] from the relations

$$F_0(Q^2) = \frac{1}{2(1+\eta)}\left[\langle +1|I^+(0)|+1\rangle + \langle 0|I^+(0)|0\rangle\right]$$

$$F_1(Q^2) = \frac{-1}{1+\eta}\sqrt{2/\eta}\langle +1|I^+(0)|0\rangle,$$
$$F_2(Q^2) = \frac{-1}{1+\eta}\langle +1|I^+(0)|-1\rangle, \qquad (4)$$

where $\eta = Q^2/4M_d^2$. The relation of these form factors to the conventional form factors $G_C, G_M$ and $G_Q$ is given in Eq. (3.33) of Ref. [14]. The light cone relations given in Eqs. (4) are equivalent with those given in Ref. [1].

The values of the form factors $G_C, G_M$ and $G_Q$ for $Q^2 = 0$, are the charge, the magnetic moment $\mu$, and the quadrupole moments $Q$, respectively. In terms of the forms factors $F_i$, they are given as

$$\mu = \frac{e}{2M_d}[2 + F_1(0)],$$
$$\frac{Q}{e} = \lim_{Q^2 \to 0} \frac{4F_2(Q^2)}{Q^2} - \frac{1}{M_d^2}[1 + F_1(0)]. \qquad (5)$$

The anomalous magnetic moment is $a = F_1(0)$. We will plot the nuclear moments of the deuteron as a function of the radius $R^2 = -6 dF(Q^2)/dQ^2|_{Q^2=0}$, where $F(Q^2)$ is chosen to be $F(Q^2) = \langle +1|I^+(0)|+1\rangle$.

The predictions for the moments can be written analytically as expectation values: [†] $F_1(0) = \langle \gamma_i \rangle = \int [d^2k]\gamma_i|\psi|^2 / \int [d^2k]|\psi|^2$ and $\lim_{Q^2 \to 0} F_2(Q^2)/Q^2 = \langle \gamma_2 \rangle$ where the $\gamma_i$ are given as

$$\gamma_1 = 2M_d(A - B) + \frac{M_d}{m_N}F_{2N}C, \qquad (6)$$
$$\gamma_2 = AB + \frac{1}{2m_N}F_{2N}BC, \qquad (7)$$

with $F_{2N} = a_p + a_n$. The factors $A, B$ and $C$ in the above equations are given by

$$A(x_i, k_{\perp i}, m_N) = \frac{1}{2\mathcal{M}} \frac{2x_2\mathcal{M}[m_N + (1-x_2)\mathcal{M}] - \vec{k}_{\perp 2}^2}{[m_N + (1-x_2)\mathcal{M}]^2 + \vec{k}_{\perp 2}^2},$$
$$B(x_i, k_{\perp i}, m_N) = \frac{1}{2(1-x_2)\mathcal{M}} \frac{2(1-x_2)x_2\mathcal{M}(m_N + x_2\mathcal{M}) - x_2\vec{k}_{\perp 2}^2}{(m_N + x_2\mathcal{M})^2 + \vec{k}_{\perp 2}^2},$$
$$C(x_i, k_{\perp i}, m_N) = \frac{[m_N + (1-x_2)\mathcal{M}]^2}{[m_N + (1-x_2)\mathcal{M}]^2 + \vec{k}_{\perp 2}^2}.$$

---

[†] $[d^2k] = d\vec{k}_1 d\vec{k}_2 \delta(\vec{k}_1 + \vec{k}_2)$. The third component of $\vec{k}$ is defined as $k_{3i} = \frac{1}{2}(x_i\mathcal{M} - \frac{m_N^2 + \vec{k}_{\perp i}^2}{x_i\mathcal{M}})$. This measure differs from the usual one used in Ref. [17] by the Jacobian $\prod \frac{dk_{3i}}{dx_i}$ which can be absorbed into the wavefunction.

We now take a closer look at the limits $R \to \infty$ and $R \to 0$. In the nonrelativistic limit we let $\beta \to 0$ and keep the nucleon mass $m_N$ and deuteron mass $M_d$ fixed. In this limit the deuteron radius $R \to \infty$ and

$$\mu_d = \frac{e}{2M_d}\left(2 + \frac{M_d}{m_N}F_{2N}\right) = \mu_p + \mu_n + \frac{2m_N - M_d}{M_d}\mu_N = -0.24\mu_N, \qquad (8)$$

with $\mu_N$ being the nuclear magneton. This result agrees with Ref. [14] and differs from the standard nonrelativistic result by the small last term. The nonrelativistic limit for the quadrupole moment is

$$\frac{Q_d}{e} = \frac{M_d - 2m_N}{2M_d m_N^2}(1 + F_{2N}) = -0.0006 \text{ GeV}^{-2}. \qquad (9)$$

To obtain the ultra-relativistic limit we let $\beta \to \infty$, while keeping $m_N$ fixed. In this limit the deuteron becomes pointlike ($R \to 0$) and the internal transverse momenta $<k_\perp^2> \to \infty$. The anomalous magnetic moment of the deuteron vanishes in this limit as can be seen in Fig. 1. The quadrupole moment takes on the canonical value $Q_d = -e/M_d^2$ as can be seen in Fig. 2.

The values for the moments in the ultra-relativistic limit can also be understood from general principles [1]. As shown in [2], the anomalous moments of a spin-one system satisfy a sum rule ($Q_d^a = Q_d + e/M_d^2$)

$$a_d^2 + \frac{2t}{M_d^2}\left(a_d + \frac{M_d}{2}Q_d^a\right)^2 = \frac{1}{4\pi}\int_{\nu_{th}^2}^{\infty}\frac{d\nu^2}{(\nu - t/4)^3}\left(\text{Im}f_P(\nu,t) - \text{Im}f_A(\nu,t)\right). \qquad (10)$$

Here $f_{P(A)}(\nu,t)$ is the non-forward Compton amplitude for incident parallel (anti-parallel) photon-deuteron helicities. This result can be understood as a generalization of the Drell-Hearn-Gerasimov sum rule to non-forward momentum transfers. Thus, in the pointlike limit where the threshold for particle excitation $\nu_{th} \to \infty$, the deuteron acquires the same electromagnetic moments $Q_d^a \to 0, a_d \to 0$ as that of the $W$ in the standard model. The approach to zero anomalous magnetic and anomalous quadrupole moments for $M_d R_d \to 0$ is shown in Figs. 1 and 2. Thus, even if the deuteron has no D-wave component, a non-zero quadrupole moment arises from the relativistic recoil correction.

A similar analysis can be performed for the axial-vector coupling. The coupling $g_A$ is given by the spin-conserving axial current $I_A^+$ matrix element

$$g_A(0) = \langle +1|I_A^+(0)|+1\rangle. \qquad (11)$$

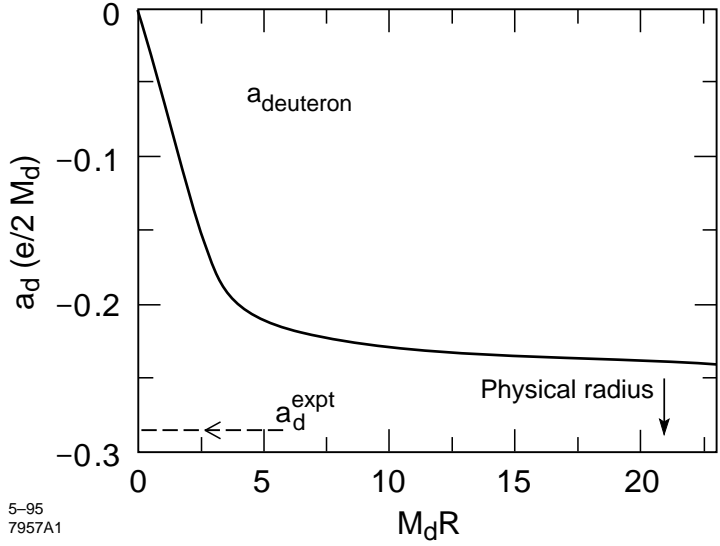

Figure 1: The anomalous magnetic moment of the deuteron as a function of its radius $R$ in Compton units. The experimental value is given by the dashed lines. The discrepancy is due to the neglect of the D-wave in the deuteron wavefunction.

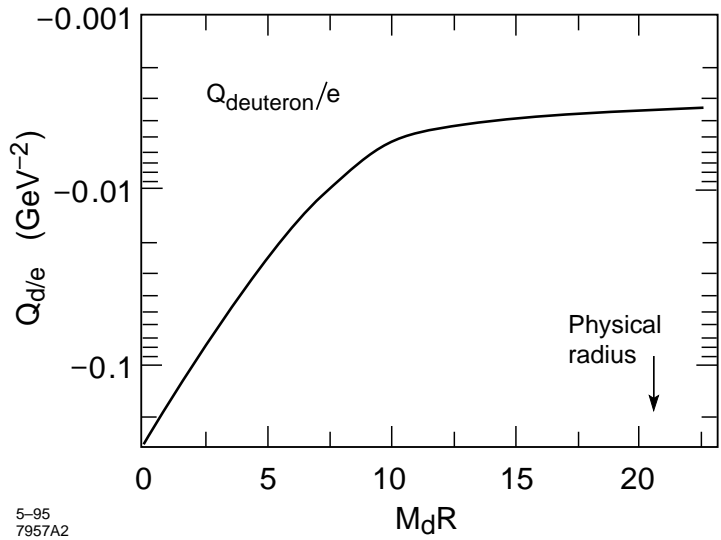

Figure 2: The quadrupole moment of the deuteron as function of $M_d R$. The quadrupole moment $Q_d \to -e/M_d^2$ for $R \to 0$.

The value for $g_A$ can be written as $g_A = \langle \gamma_A \rangle g_A^{\text{NR}}$ with $g_A^{\text{NR}}$ being the non-relativistic value of $g_A$ and with $\gamma_A$ as

$$\gamma_A(x_i, k_{\perp i}, m_N) = \frac{(m_N + x_2 \mathcal{M})^2 - \vec{k}_{\perp 2}^2}{(m_N + x_2 \mathcal{M})^2 + \vec{k}_{\perp 2}^2}. \tag{12}$$

The fact that both the axial coupling and lowest moment of the $g_1$ structure function of a composite system are modified by the Melosh transformation was first pointed out by Bucella, et al. [18], Le Youanc, et al. [19], and Close [20]. In Fig. 3 the axial-vector coupling is plotted against the deuteron radius $M_d R$. At the physical deuteron radius $M_d R = 20.6$, one predicts the value $\langle \gamma_A \rangle = 0.999$. We emphasize that at small deuteron radius the light-cone model predicts not only a vanishing anomalous moment but also

$$\lim_{R \to 0} g_A(M_d R) = 0. \tag{13}$$

As shown by Ma and Zhang [21] the Melosh rotation generated by the internal transverse momentum spoils the usual identification of the $\gamma^+ \gamma_5$ quark current matrix element with the total rest-frame spin projection $s_z$, thus resulting in a reduction of $g_A$. One can understand this physically: in the zero radius limit the internal transverse momenta become infinite and the nucleon helicities become completely disoriented.

A related quantity, the lowest moment $\Gamma_1^d = \int dx g_1^d(x)$ of the deuteron spin structure function receives the same non-additive correction $\langle \gamma_A \rangle$ as the axial-vector coupling [20]. We have to write

$$\Gamma_1^d = \frac{1}{2} (\Gamma_1^n + \Gamma_1^p) \left(1 - \frac{3}{2} \mathcal{P}_D\right) \langle \gamma_A \rangle, \tag{14}$$

where $\Gamma_1^p$ and $\Gamma_1^n$ are the moments of the nucleon structure functions, and $\mathcal{P}_D$ is the probability for the D-wave in the deuteron, which varies in the range 3-6% [22]. The relativistic binding effect in Eq. (14) can be neglected since $\langle \gamma_A \rangle = 0.999$.

In the case of the triton, the value of the Gamow-Teller matrix element is reduced by the same factor $\langle \gamma_A \rangle$ given in Eq. (12). The expectation value $\langle \gamma_A \rangle$ is for the three body case evaluated as

$$\langle \gamma_A \rangle = \frac{\int [d^3 k] \gamma_A |\psi|^2}{\int [d^3 k] |\psi|^2}. \tag{15}$$

The correction to the nonrelativistic limit for the S-wave contribution is $g_A = \langle \gamma_A \rangle g_A^{\text{NR}}$. For the physical quantities of the triton we get $\langle \gamma_A \rangle = 0.99$. This means that even at the physical radius $M_t R = 17.8$, we find a nontrivial nonzero correction of order $-1\%$ for the Gamow-Teller matrix element. This correction also characterizes the magnitude of the correction to the assumption of nucleon additivity when testing the Bjorken and Ellis-Jaffe sum rules for the helicity-dependent deep inelastic scattering structure function on He$_3$ and H$_3$ targets. Figure 4 shows that the Gamow-Teller matrix element of the triton approaches zero in the limit of small nuclear radius, just as in the case of the nucleon as a bound state of three quarks [7].

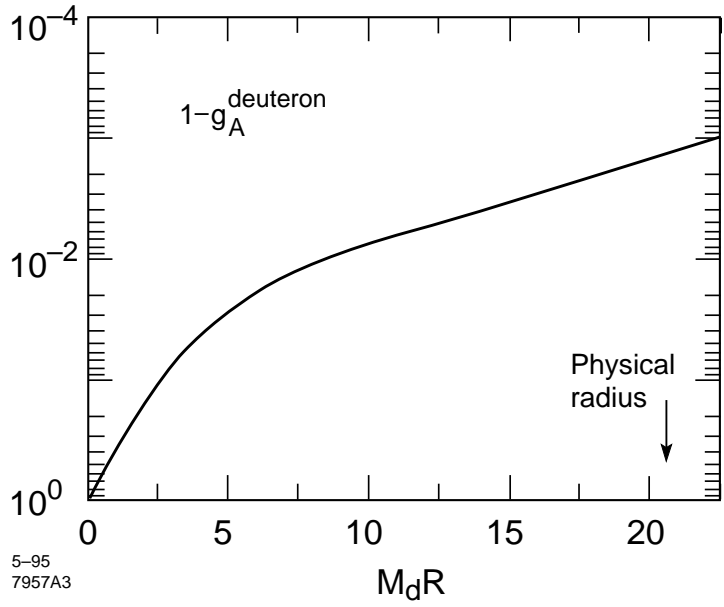

Figure 3: The axial-vector coupling of the deuteron as a function of $M_d R$.

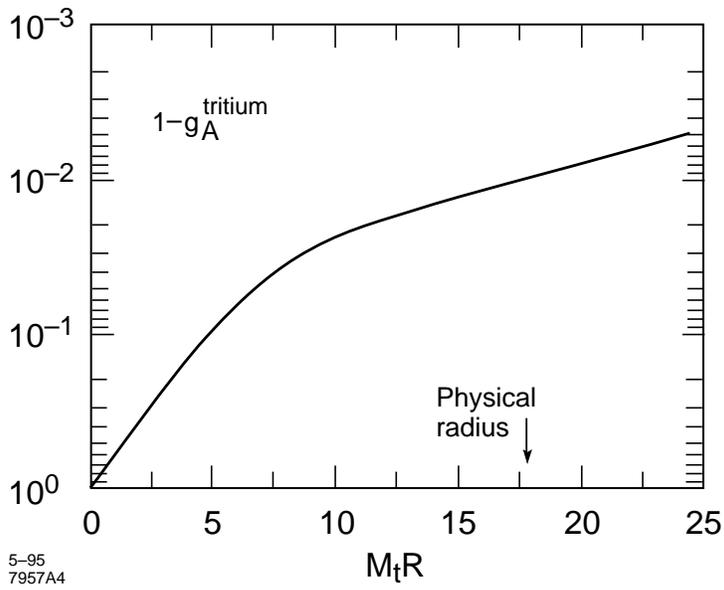

Figure 4: The axial-vector coupling (reduced Gamow-Teller matrix element) for the triton decay as a function of $M_t R$.

Although the magnetic and quadrupole moments of a composite systems are usually regarded as "static" quantities, they actually require the evaluation of the current matrix elements $< p|j_\mu|p+q >$ which are, respectively, linear and quadratic in the momentum transfer $q$. The contribution to the current matrix elements which are generated by the Wigner boost of the state from its rest frame by itself gives the Dirac contribution $\mu = eS/M$ for systems of spin $S$ and the Standard Model quadrupole moment $Q = -e/M^2$ for spin-one states. The kinematical boost contribution can be neglected compared to the dynamical contributions from light constituents $\mu = \mathcal{O}(e/m)$ or internal structure $\mu = \mathcal{O}(eR)$ and $Q = \mathcal{O}(eR^2)$ if $M/m \gg 1$ and $MR \gg 1$. Thus the usual formulas for computing moments from the sum of constituent moments is only strictly valid in the cases of systems such as atoms where the electron mass is small compared to the atomic mass and the Bohr size $R$ is large compared to the Compton scale $1/M$ of the atom. In the case of a nucleon considered as a bound state of three quarks, the relativistic effects reduce the anomalous magnetic moment and axial coupling by a factor of $\simeq 0.75$ [7, 20]. The deuteron and triton are non-relativistic bound state systems; nevertheless, we have found nontrivial finite binding corrections to the standard treatment of their magnetic and quadrupole moments.

An important consistency check of any bound state formalism is the demonstration that the magnetic and quadrupole moments of a spin-one composite system reproduces the canonical Standard Model values in the point-like limit $MR \to 0$. We have shown that the light-cone analysis correctly reproduces the correct ultra-relativistic limit for the electromagnetic moments. In addition we have shown that the axial couplings of composite spin-one systems vanish in the point-like limit. In the Standard model the parity-violating Gamow-Teller axial couplings of the $W$ and $Z$ vanish at tree level. Thus, even though composite spin-one systems are not gauge fields, their couplings can simulate the canonical axial and electromagnetic moments of the Standard Model provided they are sufficiently compact. This is interesting from the phenomenological point of view, since it keeps open the possibility that the $Z$ and $W$ vector bosons of the Standard Model could be composite provided their internal scale is sufficiently small and their excitation energies sufficiently high [23]. On the other hand, the light-cone Fock state description predicts $g_A \to 0$ for composite spin-$\frac{1}{2}$ systems in the point-like limit, whereas the canonical axial coupling in the Standard Model is $g_A = 1$ for elementary spin-$\frac{1}{2}$ fields. It thus remains an open question whether a consistent dynamical model of composite leptons and quarks [24] can be formulated which can simultaneously simulate their observed magnetic moment and axial couplings.

## Acknowledgements

We thank Frank Close, John Hiller, Eric Sather, and Mark Strikman for helpful discussions.